\documentclass[namedreferences]{solarphysics}
\usepackage[hyperref,optionalrh,solaromanenum]{spr-sola-addons} 
\usepackage{graphicx}                    
\usepackage{color}                       
\usepackage{breakurl}                         

\usepackage{array}
\newcolumntype{L}[1]{>{\raggedright\let\newline\\\arraybackslash\hspace{0pt}}m{#1}}
\newcolumntype{C}[1]{>{\centering\let\newline\\\arraybackslash\hspace{0pt}}m{#1}}
\newcolumntype{R}[1]{>{\raggedleft\let\newline\\\arraybackslash\hspace{0pt}}m{#1}}

\chardef\us=`\_

\begin{document}

\begin{article}

\begin{opening}
 
\title{Progress in Solar Cycle Predictions: Sunspot Cycles 24--25 in Perspective}


\author[addressref={aff1,aff2},corref,email={dnandi@iiserkol.ac.in}]{\inits{Dibyendu Nandy}\fnm{Dibyendu Nandy}}


\address[id=aff1]{Center of Excellence in Space Sciences India, Indian Institute of Science Education and Research Kolkata, Mohanpur 741246, West Bengal, India}
\address[id=aff2]{Department of Physical Sciences, Indian Institute of Science Education and Research Kolkata, Mohanpur 741246, West Bengal, India}

\runningauthor{Nandy et al.}
\runningtitle{Progress in Solar Cycle Predictions}

\begin{abstract}

The dynamic activity of the Sun -- sustained by a magnetohydrodynamic dynamo mechanism working in its interior -- modulates the electromagnetic, particulate and radiative environment in space. While solar activity variations on short timescale create space weather, slow long-term modulation forms the basis of space climate. Space weather impacts diverse space-reliant technologies while space climate influences planetary atmospheres and climate. Having prior knowledge of the Sun's activity is important in these contexts. However, forecasting solar-stellar magnetic activity has remained an outstanding challenge. In this review, predictions for sunspot cycle 24 and the upcoming cycle 25 are summarized, and critically assessed. The analysis demonstrates that while predictions based on diverse techniques disagree across solar cycles 24--25, physics-based predictions for solar cycle 25 have converged and indicates a weak sunspot cycle 25. It is argued that this convergence in physics-based predictions is indicative of progress in the fundamental understanding of solar cycle predictability. Based on this understanding, resolutions to several outstanding questions related to solar cycle predictions are discussed.  

\end{abstract}

%

\end{opening}


\section{The Case for Solar Cycle Predictions}

The space environment in the solar system is governed by the variable activity of the Sun. This variability is manifested in changing flux of solar radiation, solar energetic particles, solar magnetic fields and a variable solar wind output. Occasionally, energetic events such as flares and coronal mass ejections (CMEs) introduce extreme perturbations in our space environment. These phenomena are collectively referred to as space weather.
Severe space weather can impact the health of satellites and astronauts in outer space, disrupt satellite-based communications and navigational networks, high frequency radio communications, electric power grids, oil pipelines and air-traffic on polar routes. Understanding, assessing and predicting space weather is therefore critical to protection of modern day technologies and is considered a high priority research area \citep{NRC1997, NRC2013, NSTC2019, EU-JRC2016, UNOOSA2017, Schrijver2015}.

Slower, longer term modulation in the solar activity output over time scales ranging from decades to centuries to millennia \citep{Solanki2004, Usoskin2017} define what is known as space climate \citep{Versteegh2005}. Space climate plays a role in the forcing of planetary atmospheres, e.g., in the heating of the upper atmosphere and its expansion which is relevant for satellite drag and mission life-time estimates. While magnetically modulated variations in the solar irradiance provide a link to planetary climate systems \citep{Solanki2003}, solar open flux variations determine the flux of galactic cosmic rays at Earth \citep{Usoskin2002}. Secular variations in the Sun, solar wind and interplanetary magnetic flux also impacts planetary magnetospheres with consequences for geomagnetic activity \citep{Mursula2003} and atmospheric evolution \citep{Das2019}. Indeed, the intimate relationship between solar-stellar activity and the planets that they host can extend over their coupled lifetimes and is based on causal connections between physical processes in stellar interiors and planetary atmospheres \citep{Nandy2007}.

Stripped bare to its roots, space weather and space climate are fundamentally products of the solar magnetic cycle and its diverse manifestations -- which are consequences of the emergence, evolution and dynamics of solar magnetic fields or sunspots and their impact on the heliospheric environment. Thus, the quest to assess and forecast our space environment in intimately related to, and contingent upon understanding the physics of the solar magnetic cycle, and develop predictive capabilities based on this understanding.

Solar magnetic fields are generated by a magnetohydrodynamic (MHD) dynamo mechanism that is sustained by complex interactions between plasma flows and magnetic fields in the Sun's convection zone \citep{Parker1955a, Babcock1961, Leighton1969, Charbonneau2020}. The toroidal component of the dynamo generated magnetic field buoyantly emerges through the solar surface creating sunspots -- strongly magnetized, and relatively darker regions on the solar surface. Sunspots have been monitored for over four centuries starting with the pioneering observations of Galileo Galilei. Their magnetic nature was discovered in the early $20^{th}$ century \citep{Hale1908}. Development of the magnetograph instrument \citep{Babcock1955} allowed observations of the large-scale (relatively weaker) magnetic field that exists outside of sunspots and which plays a crucial role in the build up of the polar flux leading to the reversal of the global dipolar field (i.e., the poloidal component of the dynamo generated magnetic field).  

These long term observations illuminate various facets of the sunspot cycle on the one hand \citep{Hathaway2015} and on the other hand, provide important constraints on the solar dynamo mechanism \citep{Nandy2002} and motivate various approaches for solar cycle predictions \citep{Petrovay2020}.

\section{Sunspot Cycle Observations: The Prediction Challenge}

\begin{figure}[h] 
 \centerline{\includegraphics[width=1.0\textwidth,clip=]{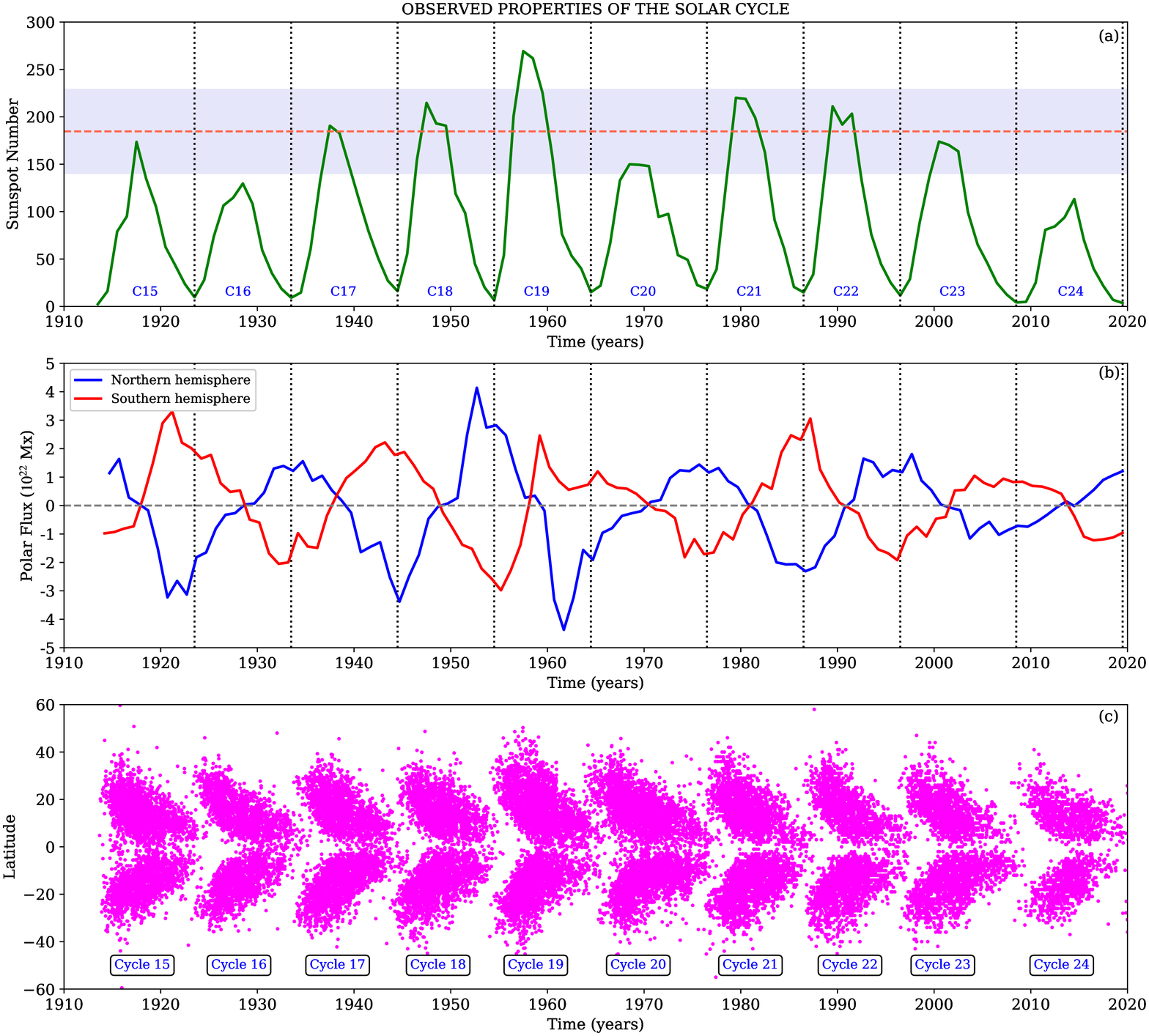}}
\caption{Solar cycle observations. The solid green curve in Figure~1a depicts the sunspot time series from 1914.5 to 2019.5. This time series is generated utilizing the revised version of the annually averaged new sunspot number data \citep{Clette2015} acquired from the World Data Center SILSO, Royal Observatory of Belgium. The red-dashed line denotes the mean (184.630) of all cycle amplitudes during this period and the shaded portion indicates 1$\sigma$ (44.282) variation around the mean. In Figure~1b, we show the evolution of the polar hemispheric flux derived from Mount Wilson Observatory calibrated polar faculae data \citep{Munoz2012} covering 1914 to 2014. This data is acquired from the solar dynamo database maintained by Andrés Muñoz-Jaramillo. We extend this plot with the calibrated Wilcox Solar Observatory polar field data till 2019.5. The blue (red) curve represents the polar flux in northern (southern) hemisphere. In Figure~1c we plot the sunspot butterfly diagram. The data for the butterfly diagram is acquired from the Royal Greenwich Observatory/USAF-NOAA active region database compiled by David H. Hathaway for the period 1914 to 2016. The subsequent data until 2019 is acquired from the Helioseismic and Magnetic Imager (HMI) instrument \citep{Scherrer2012} on board the Solar Dynamics Observatory (SDO; \cite{Pesnell2012b}). Solar cycle numbers are indicated in the top and bottom panels.}\label{fig:1}
\end{figure}

Long term observations indicate that the number of sunspots on the solar surface (which is a proxy for the toroidal component of the solar magnetic field) increases and decreases in a cyclic fashion with an average periodicity of 11 years. Barring episodes of grand minima in activity, e.g., the Maunder minimum, this trend has been maintained over the last four centuries. In Figure~\ref{fig:1} we present an overview of solar cycle observations over the last 100 years and establish century-scale solar cycle climatological trends of relevance to cycle predictions. Figure~\ref{fig:1}a presents the (revised) sunspot number time series covering cycles 15--24. It is evident that while the solar cycle period varies only slightly from cycle to cycle, there is significant variability in its amplitude -- quantified by the (annual averaged) peak sunspot number. From these observations we establish a century-scale mean sunspot cycle amplitude 184.630 $\pm$ 44.282 ($\sigma$, i.e., standard deviation). The climatological mean is indicated by the red-dashed line, while the range (mean $\pm$ 1$\sigma$) is indicated by the shaded region in Figure~\ref{fig:1}a. We define cycles whose peak lies within the shaded region (mean $\pm$ 1$\sigma$) as moderate solar cycles; cycles which lie within this region but are higher than the mean may be further sub-classified in to moderate-strong cycles and which lie below the mean may be sub-classified in to moderate-weak cycles. Extreme solar cycles which lie over this region (greater than mean $+$ 1$\sigma$) are classified as strong cycles and cycles which lie below this range (mean $-$ 1$\sigma$) are classified as weak cycles. We note that only one cycle (19) has been extremely strong, while two cycles (16 and 24) have been extremely weak. In fact the recently concluded solar cycle 24 has been the weakest cycle of the past century. There is no discernible pattern in amplitude variability from one cycle to another in the sunspot time series which makes their prediction a challenging task.  

The variation of the Sun's polar flux, which is a proxy for the poloidal component of the solar magnetic cycle, is depicted in Figure~\ref{fig:1}b. The (radial) polar flux in the solar north and south poles are found to be opposite to each other indicative of a global dipole field configuration near solar minima. The polar fields also undergo cyclic reversals. They are the weakest and reverse their sign (polarity) during sunspot maxima and they are the strongest during sunspot cycle minima; the polar field lags the sunspot cycle (i.e., the toroidal field component) with a phase difference of $90^\circ$. In fact, although not evident here, the relative orientation of bipolar sunspot pairs reverse orientation from one cycle to another, indicating that the toroidal component of the solar magnetic field also reverses from one sunspot cycle to another. 

A pattern emerges when one compares the amplitude of the polar field (Figure~\ref{fig:1}b) at sunspot cycle minima with the strength of the following sunspot cycle (Figure~\ref{fig:1}a). A stronger polar field at solar minimum is indicative of a stronger (upcoming) sunspot cycle. There is a causal basis for this connection, as the solar polar field acts as the seed which is further amplified by the Sun's differential rotation to produce the next sunspot cycle. This causal connection is the basis of one of the more successful empirical prediction techniques -- the precursor method.

Figure~\ref{fig:1}c depicts the solar butterfly diagram, which indicates that there is also a spatio-temporal pattern in the sunspot cycle. Cycles begin with sunspots appearing at mid-latitudes, with more and more spots appearing at lower and lower latitudes as the cycle progresses. This pattern is followed in both the hemispheres with the cycle eventually ending with sunspots appearing only close to the equator. This pattern repeats from one cycle to another. 

Ideally, one would expect that advances in understanding the solar dynamo mechanism and advances in methodologies for accurate solar cycle predictions would be commensurate with each other. This expectation would imply that attempts at solar cycle predictions must also imbibe many of the constraints available from solar cycle observations, and be able to explain most, if not all, features of the spatio-temporal variability in the sunspot cycle. This is the rather restrictive view that is taken in categorizing physics- or model-based predictions in this review with the additional consideration that this class of predictions must also be based on MHD models of solar magnetic field evolution. Nevertheless, several other techniques -- ranging from some which have no connection with the underlying physics whatsoever to some which draw inspiration from the underlying physics -- have been utilized for forecasting the solar cycle. In the next section we revisit such predictions for solar cycle 24 and summarize predictions made until now for solar cycle 25. These predictions are further analyzed and compared to ascertain any apparent progress over the last decade in efforts at predicting the sunspot cycle. 

\section{Predictions of Solar Cycle 24}

\begin{figure}[h] 
 \centerline{\includegraphics[width=1.0\textwidth,clip=]{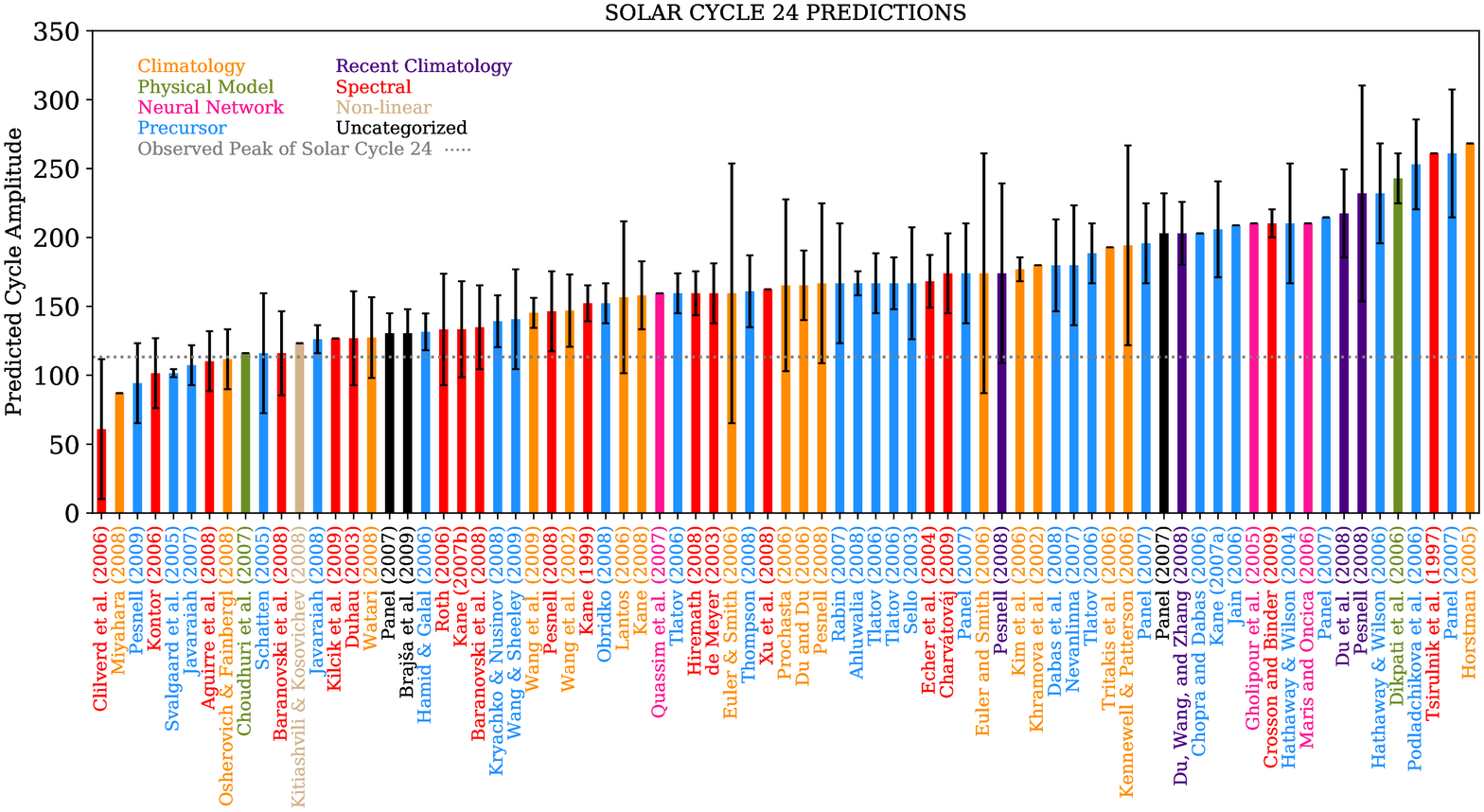}}
 \caption{Predictions of solar cycle 24 by different groups based on diverse methodologies (indicated in the plot and represented through distinct colour bars). The height of the bars indicate the predicted peak strength (scaled to conform to the new, revised sunspot time series). The mean ($\pm$ $1\sigma$) of these cycle 24 predictions is 165.390 $\pm$ 42.762 (SSN). The dashed line denotes the observed peak of solar cycle 24 (113.3 SSN in the revised scale) for comparison. Details of the utilized methodologies can be found in the references cited below the corresponding predictions; these are available in the bibliography.}\label{fig:2}
\end{figure}

Solar cycle 24 commenced following a series of sequentially weaker solar cycles and an unusually extended minimum of sunspot cycle 23 \citep{Nandy2011}. Multiple predictions were made for solar cycle 24. Following \citep{Pesnell2008, Pesnell2012a} in Figure~\ref{fig:2}, we summarize the predictions of cycle 24. In this figure, predictions have been categorized based on the underlying methodology utilized for making the forecast for the peak cycle amplitude. Note that the numbers for the predicted cycle amplitude in Figure~2 have been scaled to conform to the new, revised sunspot time series \citep{Clette2015} for ease of comparative analysis (and thus numbers are different from those in \citep{Pesnell2008, Pesnell2012a}. The observed peak amplitude of solar cycle 24 is depicted in Figure~\ref{fig:2} with a gray-dashed line.

An analysis of solar cycle 24 predictions in Figure~\ref{fig:2} reveal that a majority of the forecasts predicted a much higher cycle than what was observed. The mean ($\pm$ $1\sigma$) of all the cycle 24 predictions is 165.390 $\pm$ 42.762 in units of sunspot number (SSN). The observed peak (113.3 SSN) was outside the range of these predictions! Clearly, there was no convergence in predictions utilizing the diverse techniques. 

Evidently, there is no physical meaning in arriving at a mean or an average forecast) from such diverging predictions employing unrelated and disparate techniques; any solar cycle prediction panel should keep this in mind. Nonetheless, as we shall see, the mean and standard deviation of different cycle predictions across different cycles may provide a purely pragmatic method to assess relative consensus among techniques. 

The two physical (dynamo) model based predictions by \cite{Dikpati2006} and \cite{Choudhuri2007} predicted very strong and very weak cycles, respectively. Although the latter prediction for a weak cycle turned out to match observations, it was not immediately obvious {\it{why}}, and the non-convergence in physics-based forecasts led to massive heartburn and controversies that shook the field. To make matters worse  -- if that were possible -- the NOAA-NASA Solar Cycle Prediction Panel made an early declaration of a strong cycle and subsequently had to revise the forecast to a weak cycle after the cycle had already started! Perhaps the early panel statement was motivated from the perspective of achieving a consensus based on the many strong-cycle-forecasts and it may have been particularly influenced by the \cite{Dikpati2006} prediction.

It is natural that anyone looking at the ``confusogram'' of forecasts for solar cycle 24, their non-convergence and disagreement with the eventually observed cycle 24 peak would conclude that the understanding of solar cycle predictability was at a very immature stage at this juncture. In hindsight of solar cycle 24 and now armed with recent predictions for sunspot cycle 25 at the intervening cycle minimum, one is tempted to pose the question, are we any better off a decade down the line one solar minimum later? To assess the current scenario, we present predictions for solar cycle 25 in the next section and analyze them.   

\section{Predictions of Solar Cycle 25}

In Figure~\ref{fig:3} we present predictions of the peak amplitude of sunspot cycle 25 from various groups using a diversity of techniques. For predictions made before the year 2016, we have scaled the predicted amplitude to conform to the revised sunspot time series but have left the numbers unchanged for predictions published in the year 2016 or thereafter; i.e., we have assumed that predictions published in the year 2016 and thereafter have been calibrated with the new sunspot time series released in 2015 \citep{Clette2015}. First, we categorize solar cycle 25 predictions based on the utilized methodology and provide a brief narrative summary of each of the predictions.

\subsection{Physical Model Based Forecasts}
\begin{enumerate}
    \item\cite{bhowmik2018} utilized a observational data assimilated, century-scale calibrated SFT model whose output was coupled to a solar internal dynamo model to predict that sunspot cycle 25 would be similar or slightly stronger than solar cycle 24 with a peak SSN of 118 and a range from 109 - 155. They also predicted that the peak of solar cycle 25 would occur in 2025 ($\pm$ 1 year).
\\
    \item\cite{Jiang2018} predicted solar cycle 25 utilizing a solar Surface Flux Transport (SFT) model. They used the correlation between the axial dipole moment at cycle minimum and the subsequent cycle strength and the other empirical properties of solar cycles to predict the possible behaviours of the succeeding cycle. Their predicted peak SSN is 125 $\pm$ 32.
\\
    \item\cite{Upton2018} used their Advective Flux Transport (AFT) model and the empirical (precursor) relationship between the polar field and the subsequent cycle amplitude to predict the SSN. According to them solar cycle 25 would be slightly weaker than solar cycle 24. They have predicted a maximum sunspot number (SSN) of 110.
\\
    \item\cite{Labonville2019} used a data-driven hybrid 2 $\times$ 2D flux transport dynamo (FTD) model to forecast properties of the upcoming sunspot Cycle 25. They predicted that the maximum sunspot number SSN for solar cycle 25 would be between $89-14$ to $89+29$. The peak is predicted to occur between 2025.3+0.89 to 2025.3-1.05.
\end{enumerate}

\subsection{Precursor Technique Based Forecasts}
\begin{enumerate}

    \item\cite{HELAL2013} used a solar activity precursor technique of spotless events to predict maximum SSN of solar cycle 25 which would be 118.2. According to this study the upcoming cycle will peak between 2022-2023.
\\
    \item\cite{Pesnell2018} utilized Solar Dynamo (SODA) index that combines values of the solar polar magnetic field and the solar spectral irradiance at 10.7 cm to create a precursor of future solar activity. They predict a maximum SSN of 135 $\pm$ 25 occurring in 2025.2 $\pm$ 1.5.
\\
    \item\cite{Hawkes2018} calculated the helicity flux through both the hemispheres using a model that takes account of the Omega effect, using the magnetic field data from Wilcox Solar Observatory (WSO) covering a total of 60 years. Using various correlation analysis between helicity flux and the sunspot time series they predict the amplitude of solar cycle 25 to be 117 which is slightly higher than that of cycle 24.
\\
    \item\cite{Petrovay2018} used Rush To The Poles phenomenon (RTTP) in coronal green line emission to predict the peak SSN for solar cycle 25. Based on the correlation between the rate of the RTTP and the time delay until the maximum of the next solar cycle and the known internal regularities of the sunspot number series, they predicted that the peak amplitude to be 130 occurring in late 2024.
\\
    \item \cite{Gopalswamy2018} used the polar and low-latitude brightness temperatures as proxies for the polar magnetic field to predict cycle 25. The polar microwave brightness temperature is found to be correlated with the polar magnetic field strength and the fast solar wind speed. These correlations are used to predict a maximum SSN of 89 in the south and 59 in in the northern hemisphere (unsmoothed SSN), 116 in South and 97 in north (smoothed SSN).
\end{enumerate}

 \subsection{Non-linear Model Based Forecasts}
\begin{enumerate}
    \item \cite{Sarp2018} implemented a non-linear prediction algorithm based on delay-time and phase space reconstruction to forecast a maximum SSN of 154 $\pm$ 12 occurring in 2023.2 $\pm$ 1.1. 
\\
    \item\cite{sello2019} used revised Non Linear Dynamics methods to predict the maximum SSN for solar cycle 25 to be 107 $\pm$ 10 occurring in July 2023 $\pm$ 1 year. 
\\
    \item\cite{Kitiashvili2016} applied an ensemble Kalman filter method to predict solar cycles using a low-order, nonlinear dynamo model. They used data assimilation approach to predict a maximum SSN of 90 $\pm$ 15 occurring in 2024 $\pm$ 1 year.
    
\end{enumerate}

\subsection{Statistical Forecasts}
\begin{enumerate}
    \item\cite{Li2015} found that the ascent duration (AD) of a solar cycle is statistically related to the descent duration (DD) of the Cycle.  Statistical relations among feature parameters of the solar cycle are used to predict the behaviour of solar cycle 25. The maximum SSN is predicted to be 109.1 occurring around October 2023.
\\    
    \item\cite{Pishkalo2008} used the correlation between cycle parameters to predict SSN for solar cycle 25. According to this study the peak SSN would occur on 2023.4 $\pm$ 0.7 with an amplitude of 112.37 $\pm$ 33.4.
  \\  
  \item\cite{Li2018} utilised relations among the feature parameters of solar cycles under the bimodal distribution for the modern era cycles (10–23). These relations are utilized to predict that the solar cycle 25 would initiate in October 2020 and reach its maximum amplitude of 168.5 $\pm$ 16.3 in October 2024.
  \\
    \item\cite{Singh2017} performed a statistical test for persistence of solar activity based on the value of Hurst exponent (H). They predict that the maximum SSN would occur on June 2024 with a value of 102.8 $\pm$ 24.6.
 \\
    \item\cite{Han2019} implemented Vondrak smoothing method to produce a series of smoothed SSN (denoted SSN-VS) -- which closely mimics the 13-month running mean SSN. Applying these techniques to the descending phase of cycle 24 they make predictions for cycle 25 whose peak is estimated to be 228.8 $\pm$ 40.5 occurring in 2023.918 $\pm$ 1.64 year.
\\
    \item\cite{Du2006a} used the maximum–maximum cycle length as one of the indicators to predict the amplitude of solar cycle 25. This study found that the maximum SSN amplitude will be 102.6 $\pm$ 22.4.
\\
    \item\cite{DuNDu2006b}, based on their analysis, claims that the amplitude of a solar activity cycle is correlated with the descending time of the [$n-3$] cycle. Based on this correlation they predict a peak SSN of 111.6 $\pm$ 17.4.
\\    
    \item\cite{Hiremath2008} modeled solar cycles considering a forced and damped harmonic oscillator. They obtain long-term amplitudes, frequencies, phases and decay factors from 22 cycles (1755–1996). Using these parameters and employing a autoregressive model they predict a maximum SSN of 110 $\pm$ 11 occurring in 2023.
\\
    \item\cite{Du2006c} claim that the maximum amplitude of solar activity cycles are anti-correlated with the newly defined solar cycle lengths three cycles before. They use this correlation to predict a peak SSN of 144.3 $\pm$ 27.6.
\\
    \item\cite{Abdusamatov2007} analyzed the long-term cyclic variations of solar activity, radius, and solar constant claiming them to be correlated in both phase and amplitude. Based on this they predict a very low cycle 25 peak of 50 $\pm$ 15.
\\
\end{enumerate}

\subsection{Spectral Methods Based Forecasts}
\begin{enumerate}
    \item\cite{Kane2007b} used spectral analysis of the sunspot time series to detect periodicities by the maximum entropy method (MEM). The periodicities obtained are further utilized in a multiple regression analysis (MRA) to estimate the amplitude of cycle 25 to be between 112-127 with a mean value 119 occurring around 2022-2023.
\\
    \item\cite{Rigozo2011} decomposed monthly sunspot number data during the 1850-2007 interval (solar cycles 9–23) into several levels and searched for periodicities by iterative regression at each level. Their prediction is based on extrapolation of SSN time series spectral components. They estimate a maximum SSN of 132.1 occurring in April 2023.
\\
\end{enumerate}

\subsection{Machine Learning and Neural Network Based Forecasts} 
\begin{enumerate}
    \item \cite{Dani2019} used machine learning method of Linear Regression (LR), Random Forest (RF), Radial Basis Function (RBF) and Support Vector Machine (SVM) to predict the peak SSN for solar cycle 25. Predicted peak amplitudes are 159.4 $\pm$ 22.3, 110.2 $\pm$ 12.8, 95.5 $\pm$ 21.9 and 93.7 $\pm$ 23.2 occurring in September 2023, December 2024, December 2024 and July 2024, respectively. The predicted peak values are $114.7-23.2$ and $114.7+22.3$.
\\
    \item\cite{Quassim2007} used neuro fuzzy approach to predict solar cycle 25. According to their study the cycle maximum would have an amplitude of 116 around 2020.
\\
    \item \cite{Okoh2018} used a method known as Hybrid Regression-Neural Network that combines regression analysis and neural network learning (Ap index is used for prediction) to forecast the amplitude of solar cycle 25. They predict a peak amplitude of 122.1 $\pm$ 18.2 occurring on January 2025 $\pm$ 6.
\\
    \item \cite{Attia2013} used neural network model and found a suitable number of network inputs for the sunspot data series based on sequential forward search for the Neuro-Fuzzy model. This study predicts a peak SSN of 90.7 $\pm$ 15.
\\
    \item \cite{Covas2019} used neural network technique to perform a spatio-temporal analysis of solar cycle data and estimates the maximum SSN for cycle 25 to be 57 $\pm$ 17 occurring on 2022-2023.
\\
\end{enumerate}

\subsection{Uncategorized Forecasts}

\begin{enumerate}

    \item\cite{Javaraiah2015} studied the combined Greenwich and Solar Optical Observing Network (SOON) sunspot group data during 1874-2013. They analysed and studied the relatively long-term variations in the annual sums of the areas of sunspot groups in 0–10, 10–20, and 20–30 latitude intervals of the Sun’s northern and southern hemispheres. Long-term variations in the north–south asymmetry of solar activity is used to predict a SSN of 42 $\pm$ 13.
\\
    \item\cite{Kakad2017} estimated the Shannon entropy related to the declining phase of the preceding Solar Cycle which is used to predict SSN. Two SSN maximum for two different values of entropy are estimate at 63 $\pm$ 11.3 and 116 $\pm$ 11.3.
\\
    \item\cite{Chistyakov1983B} used regularities of secular and 22-year variations for their forecast. The predict a peak SSN of 121 occurring in 2028.5.
\\
    \item\cite{Kontor1984} utilized a hypothesis that the cycle peak envelop oscillates between the time dependent high and low levels to predict the nature of the solar cycle 25. They estimate a peak SSN of 117 around 2024.
\\
\end{enumerate}

\subsection{Analysis of Solar Cycle 25 Forecasts}

\begin{figure}[h] 
 \centerline{\includegraphics[width=1.0\textwidth,clip=]{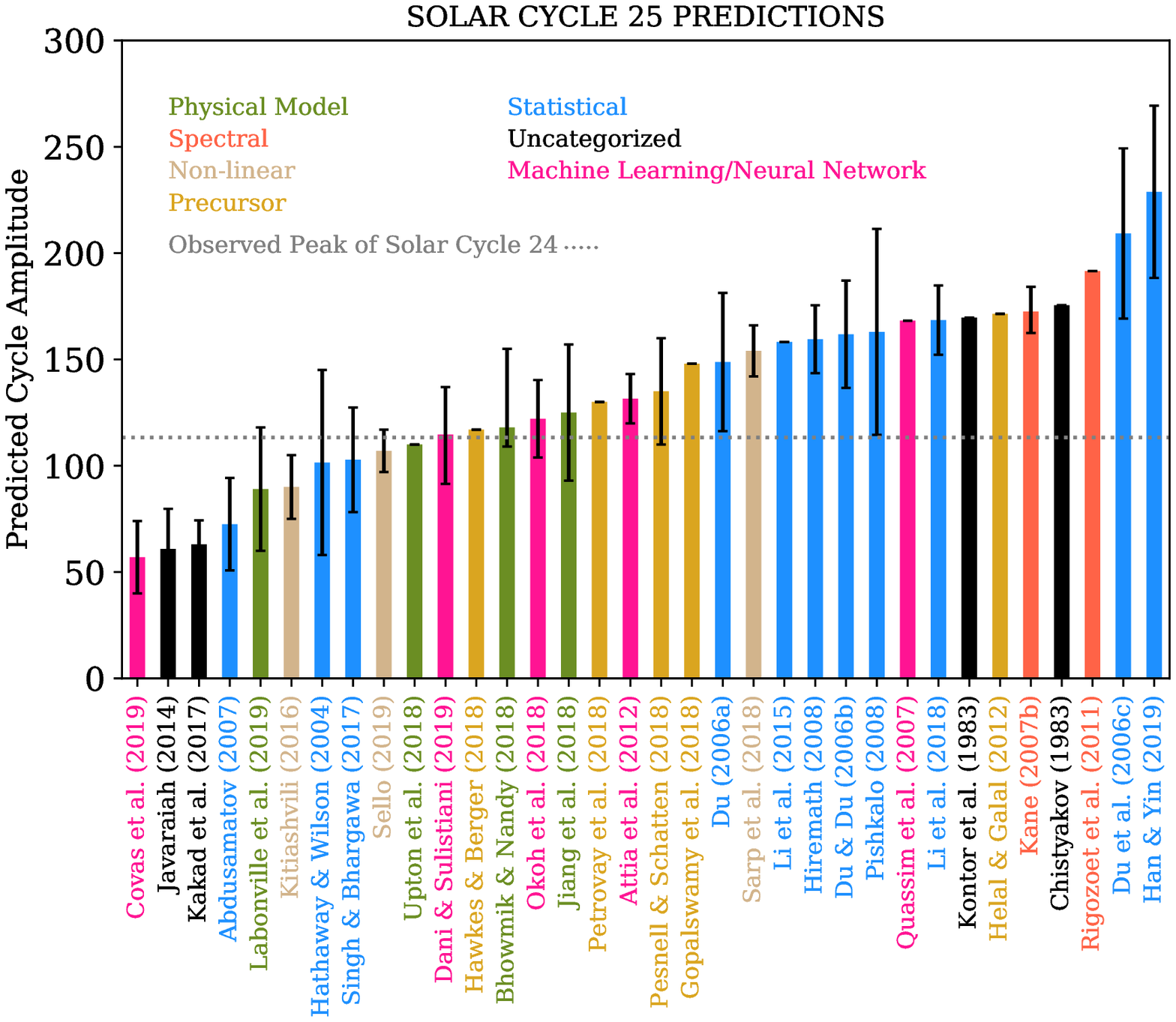}}
  \caption{Predictions of solar cycle 25 by different groups based on diverse methodologies (indicated in the plot and represented through distinct colour bars). The height of the bars indicate the predicted peak strength (scaled to conform to the new, revised sunspot time series). The mean ($\pm$ $1\sigma$) of all cycle 25 predictions is 135.88 $\pm$ 39.27 (SSN). The dashed line denotes the observed peak of solar cycle 24 (113.3 SSN in the revised scale) for comparison. Details of the utilized methodologies can be found in the references cited below the corresponding predictions; these are available in the bibliography.}\label{fig:3}
\end{figure}

In Figure~\ref{fig:3} we summarize the various predictions for solar cycle 25. For reference, the gray-dashed line indicates the observed peak amplitude of solar cycle 24 (113.3 SSN). We find that forecasts for cycle 25 based on different techniques still diverge widely, and a majority of the forecasts indicate a cycle 25 stronger than cycle 24. The mean ($\pm$ $1\sigma$) of the different predictions is 134.012 $\pm$ 39.053 which nonetheless conforms to a climatological weak cycle keeping in mind the definition of cycle strengths based on observed cycle amplitudes over the past 100 years (with reference to Figure~\ref{fig:1}). We note that there are far fewer forecasts for solar cycle 25 -- approximately half -- than there were for cycle 24; this in itself is encouraging and perhaps indicative of the realization that playing Russian roulette with solar cycle forecasting is perhaps not the best of ideas.    

On a more serious note, it is important to delve deeper in to physics-based forecasts for solar cycle 25 to ascertain whether any meaningful progress has occurred in this front. While these are part of the ``confusogram" of solar cycle 25 forecasts based on diverse techniques (Figure~\ref{fig:3}), we extract them out and analyze them separately in the next section. 

\subsection{Comparative Assessment of Physics-based Predictions of Solar Cycles 24-25}

\begin{figure}[h] 
 \centerline{\includegraphics[width=1.0\textwidth,clip=]{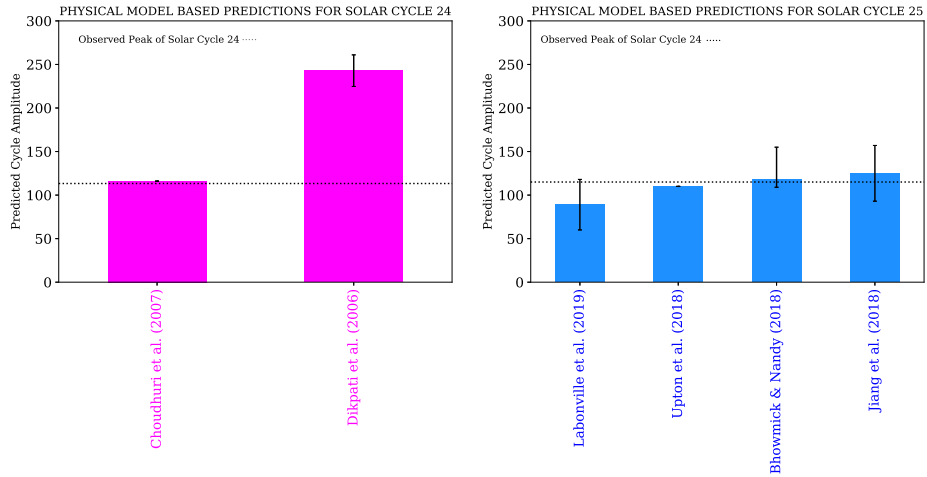}}
 \caption{Comparison of physical model based predictions for the strength of solar cycle 24 (left) and solar cycle 25 (right). For cycle 24, the mean ($\pm$ $1\sigma$) of the two physics-based predictions is 179.438 $\pm$ 63.438 (SSN). For the four physics-based predictions of solar cycle 25 the mean ($\pm$ $1\sigma$) is 110.5 $\pm$ 13.5 (SSN). The horizontal dotted line in both panels denote the observed solar cycle 24 peak amplitude (113.3 SSN) for comparison. All numbers are in the scale of the new, revised sunspot time series. Details of the utilized methodologies can be found in the references cited below the corresponding predictions.}\label{fig:4}
\end{figure}

In Figure~\ref{fig:4}, we compare those physical model based predictions of solar cycles 24 and 25 which explicitly predicted peak SSNs (as opposed to qualitative forecasts such as weak or moderate or strong cycles). We estimate the mean ($\pm$ $1\sigma$) of the physics-based predictions for cycle 24 to be 179.438 $\pm$ 63.438 (SSN). This mean is much larger than what was observed (113.3 SSN) and the standard deviation is indicative of a large divergence. 

There were four distinct physical model based forecasts for cycle 25. Two of these models \citep{Upton2018, Jiang2018} utilized different methodologies for simulating the evolution of the Sun's surface radial fields and were driven by assimilating the observed emergence profiles of bipolar active regions. They predicted the polar field expected at the end of cycle 24 and utilized the observed relationship between the polar field and the subsequent cycle amplitude (i.e., a calibrated precursor method) to forecast the peak of cycle 25. 

\cite{bhowmik2018} utilized a surface flux transport model calibrated over a century by assimilating the observed statistics of emergence of bipolar sunspot pairs to simulate the evolution of the Sun's surface radial field and polar flux. This was coupled to a solar dynamo model which assimilated the data from the surface flux transport model at every solar minima. Using this methodology they first predicted the polar field expected at the minimum of cycle 24 (in effect 4 years in advance) and subsequently utilized the century-scale dynamo simulation to forecast cycle 25. The \cite{bhowmik2018} century-scale data driven solar dynamo simulation (the first such attempt) reasonably matched past solar cycles (except the extreme cycle 19) and predicted a weak cycle 25 similar or slightly stronger than cycle 24. \citep{Labonville2019} used a slightly different methodology of coupling a surface flux transport and dynamo model more intimately. In this approach the two models communicated more frequently with each-other. \citep{Labonville2019} predicted a solar cycle 25 which is much weaker than cycle 24. 

For the four physics-based predictions of solar cycle 25 the mean predicted amplitude ($\pm$ $1\sigma$) is 110.5 $\pm$ 13.5 (SSN), i.e., very similar to the observed peak of the recently concluded solar cycle 24. 

Independently, the four physics-based forecasts for solar cycle 25 are not too distinct from each other in the sense all predict a climatogically weak sunspot cycle. More importantly, there is a small range over which the predicted uncertainties (or range of forecasts) agree. Taken together, and compared to the physics-based forecasts for cycle 24, the physics-based cycle 25 forecasts indicate significant progress towards a convergence (or a ``consensus forecast"). Is this accidental or is this convergence of predicted numbers for solar cycle 25 indicative of a convergence of fundamental ideas related to the solar dynamo mechanism? We tackle this question in the next section. 

\section{Advances in Understanding Solar Cycle Predictability}

The solar dynamo mechanism is believed to operate throughout the Sun's convection zone (and up to its surface layers) wherein, differential rotation, turbulent convection and large-scale plasma flows such as meridional circulation, turbulent flux pumping play important roles in induction and transport of magnetic fields (see, e.g., Figure~5). Following the pioneering work of \cite{Parker1955a}, there has been a community wide consensus that the toroidal component of the Sun's magnetic field is generated by the stretching of poloidal field lines by the solar differential rotation in the solar convection zone (SCZ). We note that observations and simulations of magnetic activity in other stars also bear out the importance of stellar differential rotation in sustenance of magnetic cycles \citep{Brun2015}.

The toroidal flux tubes rise up due to magnetic buoyancy \citep{Parker1955b} during which they are twisted by helical turbulent convective motions which is thought to sustain a mean-field $\alpha$-effect that can reproduce the Sun's poloidal field \citep{Parker1955a}. This process remains unobserved and unconstrained till date. Simulations of the dynamic rise of magnetic flux tubes through the SCZ show that the Coriolis force can impart a systematic tilt to bipolar sunspot pairs \citep{D'Silva1993, Fan1993} which explains the observed Joy's law for active region tilt angles \citep{Hale1919}. \cite{Babcock1961} and \cite{Leighton1969} suggested that the decay and dispersal of these tilted bipolar sunspot pairs can regenerate the Sun's large-scale poloidal (dipolar) field (Figure~5) mediated via flux transport processes providing an alternative formulation to the mean-field $\alpha$ effect. This alternative formulation came to be known as the Babcock-Leighton (hereafter BL) dynamo mechanism -- a process in which near-surface flux transport processes play a critical role in the build-up and reversal of the Sun's large-scale dipolar field (of which the observed surface radial field is a proxy). This process is observed in action on the Sun's surface. Numerous flux transport dynamo models have been built based on the Babcock-Leighton mechanism with different levels of complexity \citep{Durney1993, Durney1995, Choudhuri1995, Dikpati1999, Nandy2001, Nandy2002, Munoz_Jaramillo2010, Kumar2019} which reasonably match various solar cycle properties. Nevertheless, for long there has been no consensus on which of this two mechanisms for poloidal field generation plays an predominant role in the dynamo mechanism. The resolution to this dilemma is fundamental because the dominant poloidal source is also likely to be the primary source of variability in the solar magnetic cycle and the ability to adequately model this variability is fundamental to predictive models of the sunspot cycle. 

Careful analysis of long-term solar cycle observations relating the tilt and flux content of solar active regions (i.e. the BL source term) to the strength of the next sunspot cycle clearly implicates the BL mechanism as the primary determinant of solar cycle amplitudes \citep{Dasi_Espuig2010}. Fundamental theoretical analysis without recourse to parameterizations suggests that the surface magnetic field distribution which is a byproduct of the BL mechanism must be the primary source to the internal induction of the toroidal field \citep{Cameron2015}. Surface flux transport simulations imbibing the BL mechanism is able to reproduce the observed evolution of the Sun's large scale polar fields \citep{Jiang2014}. Coupled models of magnetic field evolution on the solar surface and in the convection zone (i.e., surface flux transport and dynamo models) successfully explain a century of solar cycle observations \citep{bhowmik2018}. These observations and theoretical simulations leave little doubt that the primary source for the Sun's poloidal field and the basis of cycle to cycle variability is the Babcock-Leighton solar dynamo mechanism driven by the emergence and dispersal of tilted bipolar sunspot pairs mediated via near-surface flows; anyone who believes otherwise is ignoring evidence -- a fundamental tenet of the scientific process.

However, both the dynamo models that were utilized for predictions of solar cycle 24 -- and whose predictions diverged widely -- were based on the BL mechanism. Does this go against the emergent understanding that the BL dynamo mechanism is the major source of variability in the solar magnetic cycle? \cite{Yeates2008} and \cite{Karak2012} demonstrated that the BL dynamo model behaves very differently under different assumed flux transport scenarios and that the relative efficacy of turbulent diffusion, meridional circulation and turbulent pumping determines the dynamical memory of the dynamo which is fundamental to predictability. \cite{Yeates2008} argued that differing assumptions related to the dominant flux transport mechanisms in the \cite{Dikpati2006} and \cite{Choudhuri2007} predictive models resulted in the discrepancy in their predictions. 

Furthermore, \cite{Yeates2008} and \cite{Karak2012} utilized stochastically fluctuating source terms in a non-linear BL dynamo model to estimate correlations between the polar field at cycle minima and subsequent cycle strengths and established that efficient transport of magnetic flux by turbulent diffusion and pumping in the SCZ reduces the dynamical memory of the sunspot cycle to only one cycle (see Figure~\ref{fig:6}). This implies that solar cycle predictions are only possible one cycle in advance, and that the polar field at cycle minima contributes only to the strength of the next sunspot cycle. 

The theoretical hypothesis of this short one cycle memory in the dynamo mechanism was soon confirmed in an analysis of the relationship between polar flux and the amplitude of different cycles by  \cite{Munoz2012}. Following their approach, we perform an analysis of the relationship between polar flux (estimated from flux calibrated faculae count) and the peak sunspot number of different cycles. This analysis, based on the revised sunspot time series is presented in Figure \ref{fig:7}. In both theoretical dynamo simulations based on the BL mechanism with fluctuating poloidal source term (Figure~6) and observed solar cycle correlations (Figure~\ref{fig:7}), we see that the toroidal flux of a cycle is not correlated with the poloidal flux measured near the poles at the end of that cycle. This indicates the the poloidal field source is stochastic (imbibing random variations) and the link of predictability is broken from the toroidal field to the poloidal field conversion process in the dynamo cycle. However, we find that these long-term statistically significant observations confirm the existence of a correlation between the polar field at the minima of a cycle [$n$] and the cycle amplitude of the next cycle [$n+1$]. This relationship is causally explained on the basis of dynamo theory, which we have established earlier, and provides the basis for predictive solar dynamo models and precursor prediction techniques -- explaining why the latter tend to be more accurate than other solar cycle prediction methods.
 
\subsection{Resolution of Outstanding Questions in Solar Cycle Predictions}

Finally, we summarize below resolutions to some outstanding questions in solar cycle predictions that were a challenge to the community about a decade back. These resolutions reflect the advances in our understanding of the physics of solar cycle predictability in the intervening period from the minimum of solar cycle 23 to the minimum of solar cycle 24 and lays the basis for future efforts in forecasting solar cycles.

\begin{enumerate}
    \item{{\bf{Is it possible to predict the sunspot cycle?}}\\
    Based on numerical simulations with stochastic and deterministic non-linear dynamo models \cite{Bushby2007} argued that the solar cycle is not predictable. Their conclusion was based on solutions to the non-linear system of equations diverging over large time-scales when slightly different initial conditions were assumed. That solutions in a non-linear dynamical systems would diverge over large timescales was well known, however, in the weather community and was demonstrated about half a century back in the pioneering work by \cite{Lorenz1963}. However, it is our considered view that \cite{Bushby2007} over-interpreted their results (or erred on the side-of-caution) to generalize their conclusion that short-term predictions are not possible. Numerous simulations with stochastically forced, non-linear dynamo models and observational analysis have since indicated that short term predictions of the upcoming solar cycle are possible (based on a causal relationship between the Sun's polar field and the toroidal field of the next sunspot cycle) \citep{Yeates2008, Dasi_Espuig2010, Karak2012, Munoz2012, bhowmik2018} have demonstrated that reasonably accurate (not exact) predictions of multiple solar cycles over a century is possible based on the current understanding. Thus, we reiterate that short-term, one cycle forecasts (at least of climatological relevance) is possible.  
    \\
    \item{\bf{What is the best proxy for solar cycle predictions?}}\\
    Theory of the solar dynamo mechanism \cite{Cameron2015}, numerical dynamo simulations \citep{Yeates2008, Karak2012, bhowmik2018}, and analysis of long-term observations \cite{Dasi_Espuig2010, Munoz2012} indicate that the best proxy for solar cycle predictions is the polar field (flux) at the minimum of the previous cycle. There is a causal relationship between the polar flux at cycle minima (which is a measure of the poloidal field strength) and the sunspot cycle amplitude (which is a measure of the underlying toroidal field) as the former acts as the source of the latter. Thus precursor technique based predictions that use direct polar field measurements or its proxy to predict the sunspot cycle are physically well founded.
    \\
    \item{\bf{How early can we predict the sunspot cycle?}}\\
    A good idea of the polar field at sunspot cycle minima is necessary to predict the next sunspot cycle. Typically, therefore, predictions made with accurate knowledge of the polar field strength at minima are likely to be more accurate. However, solar surface flux transport models can be used with synthetic input profiles of the declining phase of a cycle to predict in advance the polar field strength at the minimum of that cycle \citep{Upton2018}; this predicted polar field can be utilized in precursor methods \citep{Jiang2018} or dynamo models of the solar cycle \citep{bhowmik2018} to predict the next sunspot cycle. Such methodologies therefore can extend the prediction window of a cycle, say cycle [$n+1$]}, to a few years before the minima of cycle [$n$].
    \\
    \item{\bf{How many cycles in to the future can we predict?}}\\
    Theoretical simulations exploring the memory of the sunspot cycle based on solar dynamo simulations show that the polar field at the minima of cycle [$n$] is causally related to the toroidal field of the next, i.e., [$n+1$] cycle only \citep{Yeates2008, Karak2012} when turbulent flux transport processes dominate in the solar convection zone (see Figure~6). Long-term observations \citep{Munoz2012} of solar activity correlations confirm this (see Figure~7). Therefore, we postulate that the dynamical memory of the solar cycle -- as far as cycle to cycle variations are concerned -- is short. Reasonably accurate predictions are possible only for the next sunspot cycle, and not beyond.  
    \\
    \item{\bf{What properties of the solar cycle can we predict?}}\\
    The strength of the sunspot cycle as well as its timing can be approximately predicted. For example, \cite{bhowmik2018} predicted the complete profile of sunspot cycle 25 based on a combination of a solar surface flux transport model and a solar dynamo model. Because the strength of a sunspot cycle is related to its rate of rise, the latter can also in fact be a byproduct of solar cycle predictions. 
    \\
     \item{\bf{What physical dynamo model of the solar cycle is best suited for predictive purposes?}}\\
     Recent progress in solar dynamo theory and modeling and observational evidence together indicate that the Babcock-Leighton mechanism for poloidal field generation is the primary source of variability in the solar cycle \citep{Dasi_Espuig2010, Cameron2015,bhowmik2018}. These models can also be constrained by observations and driven by data assimilation. Philosophically and logically therefore, solar dynamo models based on the Babcock-Leighton framework should be utilized for predictive purposes.
    \\ 
     \item{\bf{Has convergence been achieved in physics-based solar cycle predictions for sunspot cycle 25?}}\\
     Yes, based on our analysis we conclude that physics-based forecasts for solar cycle 24 have converged and agree with each other with minor differences (Figure 4). These differences may result from disparate modeling techniques and (or) data assimilation methodologies. Our analysis reveals all the physics-based predictions of solar cycle 25 -- based on the Babcock-Leighton framework -- predict a climatologically weak cycle. Considering the range of uncertainty in these forecasts, it is safe to say that sunspot cycle 25 would be a weak to a moderately weak cycle that will peak around 2024 ($\pm 1$).   
    \\
    \item{\bf{With what accuracy can we predict the solar cycle?}}\\
    Some uncertainties in prediction are bound to result from the many uncertainties and parameterizations involved in modeling. Models that make early predictions are perhaps prone to larger uncertainties because of the higher probability of statistically extreme fluctuations, e.g., appearance of anomalous active regions \citep{Nagy2017} in the intervening prediction window. However, many of the physical model-based predictions can account for reasonable uncertainties through ensemble forecasts that provide a range of values for the predicted amplitude of the solar cycle \citep{bhowmik2018,Jiang2018,Labonville2019}. 
    \\
     \item{\bf{What is the best approach to achieving a consensus forecast in any Solar Cycle Prediction Panel?}}\\
     A large number of solar cycle forecasts utilizing a large number of techniques resulting in greater divergence indicates a statistical reality rather than a great scientific debate of equally viable ideas. This is well proven in our comparative analysis of predictions for solar cycles 24 and 25. The path towards consensus prediction from any solar cycle prediction panels such as the NOAA-NASA Prediction Panel is then well defined. Such panels must consider the underlying physics of solar cycle predictability and seriously assess only those methods -- and their agreements or disagreements -- which are rooted in firm physical foundations. It is clear from our analysis that while predictions for solar cycle 25 (Figure~3) utilizing diverse techniques still suffer from non-convergence just like cycle 24 (Figure~2), physically well founded model-based predictions for sunspot cycle 25 have converged (Figure~4). The basis of the consensus and any disagreements must be clearly declared for the community to understand and appreciate the subtleties involved in the prediction. Finally, the research manuscripts that have contributed to the consensus prediction must be disclosed. On the one hand, this allows independent scrutiny and analysis by the community, and on the other hand this provides due credit to the researchers whose work inform and contribute to a consensus prediction.
\end{enumerate}

\section{Concluding Remarks}\label{s:3}

In summary, here we review predictions of sunspot cycles 24-25 from different groups based on diverse techniques and perform a comparative analysis of these predictions. 

Our analysis reveals that while predictions based on diverse techniques continue to disagree across sunspot cycles 24-25, physical model based forecasts for solar cycle 25 have converged. This convergence indicates a weak to moderately weak sunspot cycle 25. We argue that this convergence in physics-based predictions is a consequence of the convergence of ideas and new insights in to the physics of solar cycle predictability. In particular, we note there is now overwhelming evidence that the Babcock-Leighton mechanism is the dominant driver of solar cycle variability and that the dynamical memory of the solar cycle is short, allowing for predictions of only the next cycle.  

Following the early disagreements and controversies related to solar cycle 24 predictions, significant progress has been achieved in the intervening decade, between the minimum of solar cycle 23 and solar cycle 24. This progress is presented and discussed in the light of resolutions to many outstanding questions related to solar cycle predictability. It is our hope that this progress will lay the foundations of more accurate, physics-based predictive models of the sunspot cycle on the one hand, and on the other hand, will more usefully constrain the fundamental physics of solar and stellar magnetic cycles.  

%
\begin{acks}
This review is dedicated to the memory of Bernard Durney who passed away last year somewhere in the South of France, his last years spent in relative obscurity far away from the solar physics community. Bernard made fundamental contributions to the development of Babcock-Leighton models of the solar cycle, including elucidating the role of meridional circulation in the near-surface evolution of the Sun's large-scale dipolar magnetic fields. I first started corresponding with him as a PhD student from India and I am indebted to him for his generosity in sharing his knowledge and debating ideas with someone he had never met. In fact, although we corresponded for many years, I never got a chance to meet him in person. I am grateful to Soumyaranjan Dash and Shaonwita Pal for assistance with literature survey and preparation of some of the figures.  I acknowledge utilization of data from the NASA/SDO HMI instrument maintained by the HMI team, the Royal Greenwich Observatory/USAF-NOAA active region database compiled by David H. Hathaway and MWO calibrated polar faculae data from the solar dynamo database maintained by Andrés Muñoz-Jaramillo. I acknowledge utilization of the hemispheric polar field data obtained by J. Todd Hoeksema and many dedicated graduate students at Stanford University's Wilcox Solar Observatory . The Wilcox Solar Observatory is currently supported by NASA. I acknowledge usage of the yearly mean sunspot number data from the Solar Influences Data Analysis Centre (SIDC) at the Royal Observatory of Belgium. Much of the understanding related to the solar magnetic cycle and its predictability has resulted from confronting theoretical dynamo models with these long-term solar activity databases and the continued sustenance of these databases cannot be overemphasized. The Center of Excellence in Space Sciences India (CESSI) is funded by the Ministry of Human Resource Development, Government of India, under the Frontier Areas of Science and Technology (FAST) scheme. Finally, I am grateful to the solar physicists of Argentina, and its wonderful people, for an inspiring time during my sabbatical visit to that country in connection to the 2019 total solar eclipse -- during which the idea and early work for this review was initiated.
\end{acks}

\bibliographystyle{spr-mp-sola}
\nocite{*}
\bibliography{ms}

%

\pagebreak

\pagebreak

\pagebreak

\begin{figure}[h] 
 \centerline{\includegraphics[width=1.0\textwidth,clip=]{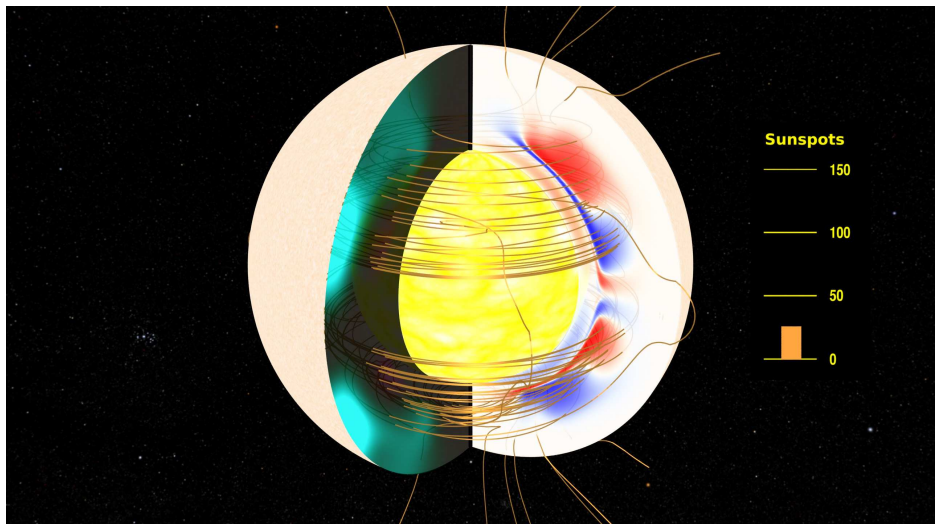}}
 \caption{Artistic rendering of a solar dynamo simulation. A cutout of the solar interior reveals the solar convection zone. The dynamo generated toroidal component of the magnetic field which forms sunspots is represented in the right-hand meridional plane (opposite polarities denoted in red and blue). The dynamo generated poloidal component of the magnetic field is represented in the left hand meridional plane (in green). The poloidal component is approximately dipolar at this phase of the simulation (near solar minimum). The right hand meridional plane shows that the toroidal field of the next cycle is already being inducted at high latitudes from this dipolar field component of the previous cycle. Credits: Simulation data by \cite{Nandy2011} and rendering by Tom Bridgman; NASA Scientific Visualization Studio.}\label{fig:5}
\end{figure}

\pagebreak

\begin{figure}[h] 
 \centerline{\includegraphics[width=1.0\textwidth,clip=]{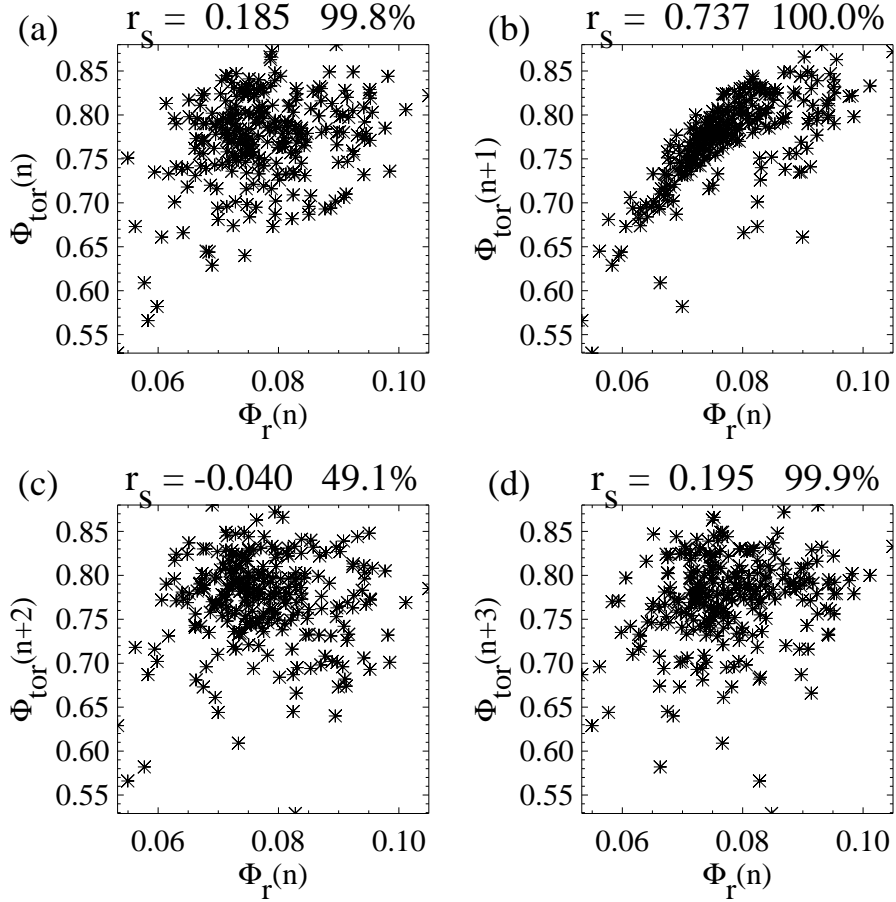}}
 \caption{Simulated cycle to cycle correlations between the polar flux ($\phi_r$) at cycle minima (say, cycle [$n$]) and the cycle amplitude ($\phi_{tor}$) of different cycles, namely (a) cycle [$n$], (b) cycle [$n+1$], (c) cycle [$n+2$] and (d) cycle [$n+3$]. This figure is reproduced from \cite{Yeates2008} and shows results of a stochastically forced, non-linear dynamo simulation based on the Babcock-Leighton mechanism. The analysis by \cite{Yeates2008} established that the solar dynamo has a short, one cycle memory, wherein, the polar field at any cycle minima contributes only to the next cycle amplitude (as evident in panel b).}\label{fig:6}
\end{figure}

\pagebreak

\begin{figure}[h] 
 \centerline{\includegraphics[width=1.0\textwidth,clip=]{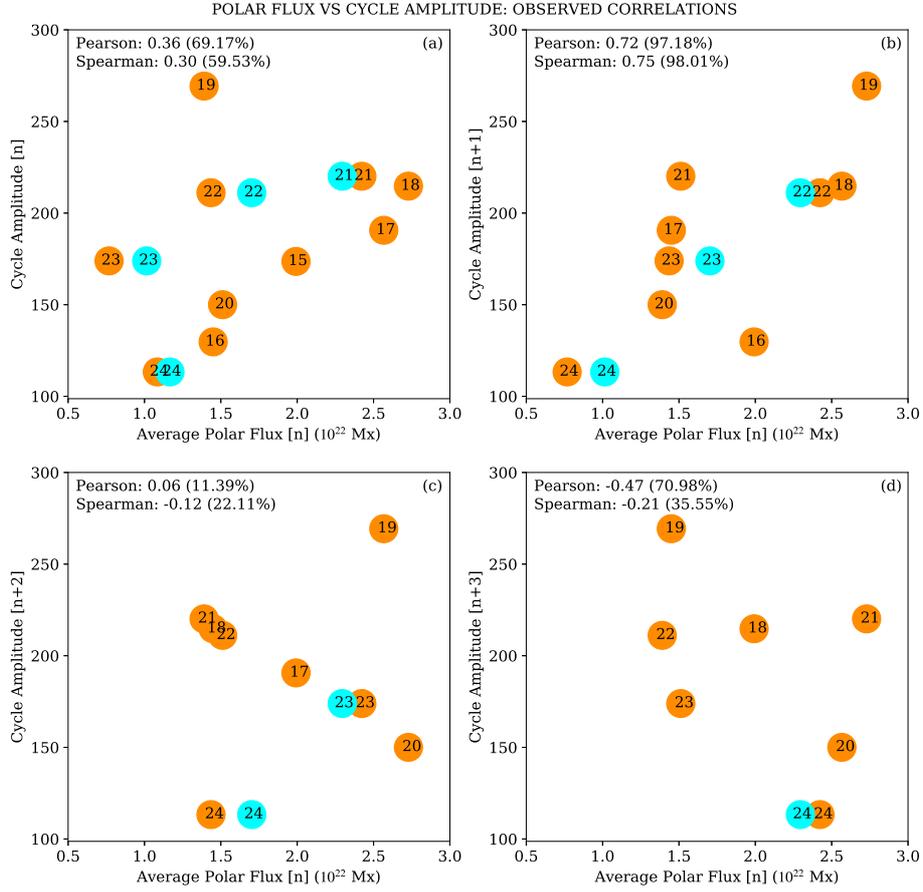}}
 \caption{Observed cycle to cycle correlations between the polar flux at cycle minima (say, $n$) and the cycle amplitude of different cycles, namely (a) cycle [$n$], (b) cycle [$n+1$], (c) cycle [$n+2$] and (d) cycle [$n+3$]. The orange filled circles represent the analysis carried out using average polar flux (derived from polar faculae) and cycle amplitude whereas the cyan filled circles show the relationship between the average dipole moment (scaled appropriately to place them in the figure) and solar cycle amplitude. The numbers inside the circles indicate the corresponding solar cycle numbers. For average dipole moment calculations we have used polar field data from the Wilcox Solar Observatory (WSO). The only significant correlation recovered is between the polar flux at the minima of a cycle, say [n] and the amplitude of the next cycle [n+1] as evident in panel (b).}\label{fig:7}
\end{figure}

\pagebreak


\begin{table}[ht]
\caption{A list of predictions for the solar cycle 25 by different groups using diverse methods.}
\label{T-complex}
\begin{tabular}{ L{3cm}  L{3cm}  L{3cm}  L{1.5cm} }
\hline
\textbf{Authors} & \textbf{Predicted SSN} & \textbf{Time} & \textbf{Category}\tabnote{S: Statistical/Correlation analysis; P: Precursor technique; MB: Model Based; N: Non-linear techniques; ML/NN: Machine Learning or Neural Network method; SP: SPectral method. } \\
\hline
\cite{Chistyakov1983B} & 121 & 2028.5 & --\\
\cite{Kontor1984} & 117 & 2024 & -- \\
\cite{Quassim2007} & 116 & 2020 & ML/NN \\
\cite{Javaraiah2015} & 42 $\pm$ 13 & -- & -- \\
\cite{Li2015} & 109.1 & Oct 2023 & S\\
\cite{Pishkalo2008} & 112.37 $\pm$ 33.4 & 2023.4 $\pm$ 0.7 & S \\
\cite{Li2018} & 168.5 $\pm$ 16.3 & Oct 2024 & S \\
\cite{Singh2017} & 102.8 $\pm$ 24.6 & June 2024 & S\\
\cite{Gopalswamy2018} & 148 & -- & -- \\
\cite{HELAL2013} & 118.2 & 2022-2023 & P \\
\cite{Pesnell2018} & 135 $\pm$ 25 & 2025.2 $\pm$ 1.5 & P\\
\cite{bhowmik2018} & 118 & 2024 $\pm$ 1 & MB\\
\cite{Labonville2019} & 89-14/89+29 & 2025.3+0.89/2025.3-1.05 & MB\\
\cite{Upton2018} & 110 & -- & MB \\
\cite{Sarp2018} & 154 $\pm$ 12 & 2023.2 $\pm$ 1.1 & N \\
\cite{Han2019} & 228.8 $\pm$ 40.5 & 2023.918 $\pm$ 1.64 & S\\
\cite{Kakad2017} & 63$\pm$11.3 or 116$\pm$11.3 & -- & -- \\
\cite{sello2019} & 107 $\pm$ 10 & July 2023 $\pm$ 1 & N \\
\cite{Okoh2018} & 122.1 $\pm$ 18.2 & January 2025 $\pm$ 6 & ML/NN\\
\cite{Du2006a} & 102.6 $\pm$ 22.4 & -- & S\\
\cite{Kane2007b} & 112 to 127 (mean 119) & 2022-2023 & SP\\
\cite{DuNDu2006b} &  111.6 $\pm$ 17.4 & -- & S\\
\cite{Attia2013} & 90.7 $\pm$ 8 & 2022 & ML/NN\\
\cite{Jiang2018} & 125 $\pm$ 32 & -- & MB \\
\cite{Covas2019}) & 57 $\pm$ 17 & 2022-2023 & ML/NN \\
\cite{Rigozo2011} & 132.1 & April 2023 & SP \\
\cite{Hawkes2018} & 117 & -- & P \\
\cite{Petrovay2018} & 130 & Late 2024 & P \\
\cite{Kitiashvili2016} & 90 $\pm$ 15 & 2024 $\pm$ 1 & N \\
\cite{Hiremath2008} & 110 $\pm$ 11 & 2023 & S\\
\cite{Dani2019}\tabnote{The predicted SSN is the mean of their four predictions using different machine learning classifiers. The errorbar is chosen to include the maximum range of the predictions.}  & 114.7-23.2/114.7+22.3 & Sep 2023 & ML/NN \\
\cite{Hathaway2004} & 70 $\pm$ 30 & 2023 & S \\
\cite{Du2006c} & 144.3 $\pm$ 27.6 & -- & S\\
\cite{Abdusamatov2007} & 50 $\pm$ 15 & -- & S \\
\hline
\end{tabular}
\end{table}

%

\end{article} 
\end{document}